\documentclass[aps,prd,singlecolumn,showpacs,superscriptaddress,longbibliography]{revtex4-2}

\usepackage{amsmath,amssymb}
\usepackage{graphicx}
\usepackage{physics}
\usepackage{hyperref}

\begin{document}

\title{Sensitivity of Standard Model Vacuum Stability to Enhanced
Scalar Couplings: A Coupling Scan and its Implications for Radiatively
Broken Electroweak Symmetry}

\author{Farrukh A.~Chishtie}
\email{fachisht@uwo.ca}
\affiliation{Peaceful Society, Science and Innovation Foundation,
Vancouver BC, Canada}
\affiliation{Department of Occupational Science and Occupational Therapy,
University of British Columbia, Vancouver BC, Canada}

\author{Sirous Homayouni}
\affiliation{Department of Mathematics, DePauw University,
Greencastle IN 46135, USA}

\date{\today}

\begin{abstract}
We study how Standard Model vacuum stability depends on the Higgs quartic
coupling at the electroweak matching scale, parameterized through a
dimensionless enhancement factor $k = \lambda_{\rm enhanced}(M_t)/
\lambda_{\rm SM}(M_t)$, with the observed Higgs mass held fixed at
$M_h = 125$~GeV. This is a scan of the coupling itself, treated as a
free parameter set by physics beyond the Standard Model, and is
distinct from the Higgs-mass scan that underlies the conventional
stability bound. Using the complete three-loop renormalization group
equations including all gauge and Yukawa couplings, we find a critical
threshold $k_{\rm crit} \approx 1.076$ separating the metastable
Standard Model trajectory from absolutely stable trajectories. At this
threshold the matching-scale coupling is $\lambda(M_t) \approx 0.135$,
which reproduces the established three-loop absolute-stability boundary,
providing an internal consistency check of the framework. The instability
scale is highly sensitive to $k$ near the threshold, with a logarithmic
susceptibility $d\ln\Lambda_I/d\ln k$ of order $10^3$ as $k \to
k_{\rm crit}^-$, larger by two to three orders of magnitude than the
analogous susceptibility of $\Lambda_{\rm QCD}$ to $\alpha_s(M_Z)$. For
$k > k_{\rm crit}$ the vacuum is absolutely stable and $\lambda$ develops
an ultraviolet Landau pole whose scale falls rapidly with increasing $k$:
for moderate enhancement the pole lies near the Planck scale, while for
the large enhancement $k \approx 7.2$ associated with radiative
electroweak symmetry breaking the matching-scale coupling is
$\lambda(M_t) \approx 0.9$ and perturbative validity is lost by
$\sim 10^5$~GeV. The radiative symmetry breaking scenario therefore points
not to grand-unified-scale dynamics but to new strong dynamics in the
$10^4$--$10^6$~GeV range, a sharp and testable prediction. The 125~GeV
Higgs mass, lying near the metastability boundary, makes the scalar
sector a sensitive probe of beyond-Standard-Model physics across the full
range of enhancements.
\end{abstract}



\maketitle

\section{Introduction}
\label{sec:intro}

The discovery of a 125~GeV Higgs boson~\cite{Aad2012,Chatrchyan2012}
confirmed electroweak symmetry breaking while revealing a fundamental
puzzle: the Standard Model (SM) Higgs quartic coupling
$\lambda(M_t) = 0.126$~\cite{Buttazzo2013} turns negative at energy scales
$\mu \sim 10^{10}$~GeV, rendering the electroweak vacuum
metastable~\cite{Degrassi2012}. While the tunneling timescale vastly
exceeds the age of the universe, this precarious proximity to the
stability boundary is theoretically unsettling and suggests deeper
physics.

The question of SM vacuum stability has attracted sustained attention
since the Higgs discovery. Buttazzo et al.~\cite{Buttazzo2013} provided
the foundational three-loop analysis characterizing near-criticality in
the $(M_h, M_t, \alpha_s)$ parameter space, locating the absolute-stability
boundary at $M_h \approx 129.6$~GeV for central $M_t$ and $\alpha_s$.
More recently, Hiller, H\"ohne, Litim, and Steudtner extended this
program systematically: in Ref.~\cite{Hiller2022} they addressed SM
metastability as a model-building task, charting the landscape of Higgs
stability using SM extensions with vectorlike fermions via gauge and
Yukawa portal mechanisms. In Ref.~\cite{Hiller2024} they revisited SM
vacuum stability using the highest available perturbative orders and
2024 PDG inputs, finding that reducing uncertainties in $M_t$ and
$\alpha_s$ by factors of two to three could establish or refute stability
at the $5\sigma$ level, and further examined singlet scalar Higgs portal
extensions to identify ``Planck-safe'' parameter spaces.

Our work addresses a complementary question. Rather than mapping the
stability boundary in $(M_h, M_t, \alpha_s)$ space or constructing
specific BSM models, we ask how the high-energy vacuum structure depends
on the value of $\lambda(M_t)$ itself, treated as a free parameter and
parameterized by a dimensionless enhancement factor $k$. We emphasize
the distinction between this coupling scan and the conventional mass
scan, since the two are easily conflated.

In the SM, the quartic coupling at the matching scale is fixed by the
physical Higgs mass through the matching relation
$\lambda(M_t) = f(M_h, M_t, \alpha_s)$, and the stability bound is
therefore naturally expressed as a condition on $M_h$. That analysis is a
\emph{mass} scan: $M_h$ is varied at fixed $M_t$, and $\lambda(M_t)$
follows through the SM relation. Here we scan a logically independent
variable. The matching-scale coupling is treated as the free parameter,
$\lambda(M_t) = k\,\lambda_{\rm SM}(M_t)$, with the observed Higgs mass
held fixed at $M_h = 125$~GeV. The enhancement $k$ represents a
beyond-Standard-Model contribution to the quartic at the matching scale
and is, by construction, decoupled from $M_h$: the relation
$\lambda(M_t) = f(M_h)$ on which the mass scan relies is precisely what a
BSM enhancement supersedes. A value of $k$ is accordingly not to be read
as a shifted Higgs pole mass, and the present scan is not a relabelling
of the $M_h$ axis. The two are complementary questions addressed through
the same SM renormalization group evolution: Ref.~\cite{Buttazzo2013}
maps stability across the SM $(M_h, M_t, \alpha_s)$ parameter space, while
the present analysis maps it across BSM enhancements of $\lambda(M_t)$ at
the observed Higgs mass.

Although the two analyses are physically distinct questions, they share
the same underlying renormalization group flow, and this provides a
stringent consistency check. The stability of a given trajectory is a
property of its boundary value $\lambda(M_t)$ under the SM RG equations,
independent of how that value is generated. We find $k_{\rm crit}
\approx 1.076$, corresponding to a matching-scale coupling
$\lambda(M_t) \approx 0.135$. This value coincides with the boundary
$\lambda(M_t)$ obtained from the conventional analysis at
$M_h \approx 129.6$~GeV, as it must, confirming that the present
three-loop implementation reproduces the established stability boundary
when expressed in the common variable $\lambda(M_t)$.

Alternative symmetry-breaking mechanisms motivate the investigation of
larger enhancements. The Coleman-Weinberg mechanism~\cite{Coleman1973}
generates masses radiatively through quantum corrections without requiring
an explicit tree-level mass term, addressing aspects of the hierarchy
problem through dimensional transmutation. Steele and
Wang~\cite{Steele2013} demonstrated that radiative electroweak symmetry
breaking (RBEWS) remains consistent with the observed 125~GeV Higgs mass.
Their analysis using Pad\'e approximation methods to nine-loop order in
the Coleman-Weinberg (CW) renormalization scheme predicts a scalar
coupling enhanced by a factor $e_{125} \approx 7.2$ compared to
conventional spontaneous symmetry breaking at the electroweak scale
within that scheme. A grand-unified realization of an enhanced quartic,
within the SO(10) framework with threshold corrections from heavy gauge
and Higgs multiplets, has been developed in Ref.~\cite{Chishtie2026PLB},
with a calculated matching-scale value $k(M_t) \approx 6.49$ consistent
with current ATLAS bounds on the Higgs self-coupling.

In this paper we perform a comprehensive analysis of SM vacuum stability
as a function of the enhancement factor $k$, using the complete
three-loop SM RG equations~\cite{Buttazzo2013}, mapping the phase
structure for $k$ from unity (SM baseline) to 7.2 (RBEWS motivation).
The threshold $k_{\rm crit} \approx 1.076$, an enhancement of roughly
$7.6\%$, separates metastability from absolute stability. The instability
scale responds steeply to $k$ near the threshold, a logarithmic
susceptibility of order $10^3$. For $k > k_{\rm crit}$, the coupling
remains positive at all scales and grows toward a Landau pole; crucially,
the pole scale is not fixed near the grand-unified scale but falls rapidly
with increasing $k$, reaching $\sim 10^5$~GeV already for $k \approx 7.2$.
This collapse of the breakdown scale is the central revision relative to
naive expectations and reshapes the physical interpretation of the
large-enhancement regime, as discussed in Sec.~\ref{sec:validity}.

The paper is organized as follows. Section~\ref{sec:rg} presents the
RG framework and numerical implementation. Section~\ref{sec:results}
presents the phase structure and quantitative results.
Section~\ref{sec:discussion} discusses implications, validity, and
connections to the recent literature. Section~\ref{sec:conclusions}
summarizes our conclusions.

\section{Renormalization Group Analysis}
\label{sec:rg}

\subsection{Standard Model Baseline}

In the SM with potential $V = -\frac{1}{2}m^2|H|^2 + \lambda|H|^4$,
precision calculations~\cite{Buttazzo2013} establish the quartic coupling
at the top mass scale:
\begin{equation}
\lambda_{\rm SM}(M_t) = 0.12604 \pm 0.00030
\end{equation}
where $M_t = 173.34$~GeV. Three-loop RG evolution drives this coupling
negative at $\Lambda_I \approx 10^{10}$~GeV, indicating vacuum
metastability with tunneling rate
$\Gamma/V \sim 10^{-600}$~yr$^{-1}$~\cite{Degrassi2012}, rendering
the instability cosmologically irrelevant but theoretically troubling.
The near-criticality of the SM has been extensively
studied~\cite{Buttazzo2013,Hiller2022,Hiller2024}, with the conclusion
that precision measurements of $M_t$ and $\alpha_s$ are the key
experimental handles within the SM context.

\subsection{Enhancement Factor Parameterization and the Coupling Scan}

We adopt a parameterization of enhanced scenarios through the
dimensionless factor:
\begin{equation}
k = \frac{\lambda_{\rm enhanced}(M_t)}{\lambda_{\rm SM}(M_t)}
\label{eq:k}
\end{equation}
This isolates the physics of sensitivity from technical scheme-dependence
issues and is independent of any specific BSM construction at the level
of the electroweak matching condition. The parameter $k$ characterizes
the ratio of the quartic coupling in an enhanced scenario relative to
the SM value, both evaluated at the top mass scale $M_t$ in the
$\overline{\rm MS}$ scheme.

As emphasized in Sec.~\ref{sec:intro}, the scan over $k$ is a scan of
the coupling itself, with $M_h$ held fixed at the observed value. It is
not a relabelling of the Higgs-mass axis, and a value of $k$ is not to
be reinterpreted as a shifted pole mass through the SM matching relation,
which a BSM enhancement supersedes. The two analyses share the same SM
renormalization group flow, so the critical value of $\lambda(M_t)$ that
separates metastability from absolute stability is common to both; what
differs is the physical question each addresses and the parameter through
which the boundary is reached.

For RBEWS, Steele and Wang's prediction~\cite{Steele2013} of
$e_{125} \approx 7.2$ at the electroweak scale motivates investigation
of $k \sim 7$. We emphasize that $e_{125}$ is defined within the CW
renormalization scheme as the ratio
$\lambda_{\rm RBEWS}(v)/\lambda_{\rm CSB}(v) \approx 0.23/0.032$,
i.e., the ratio of the radiatively generated coupling to the conventional
symmetry breaking value in the same CW scheme, both at the vev scale $v$.
This ratio provides the motivating benchmark value for $k$ at $M_t$,
with $k$ entering our analysis as an initial condition for the
$\overline{\rm MS}$ coupling. The sensitivity analysis and critical
threshold we identify are independent of the specific RBEWS value
$k = 7.2$; they hold for all $k > k_{\rm crit}$.

\subsection{Three-Loop Beta Functions and Role of Yukawa Couplings}
\label{sec:beta}

We employ the complete three-loop SM RG equations from
Ref.~\cite{Buttazzo2013}, which include all gauge couplings, the
top Yukawa coupling $y_t$, and their full interplay. Our analysis uses
the full Standard Model renormalization group equations throughout,
not those of any simplified scalar theory. The Yukawa coupling effects,
particularly the negative top quark contribution $-3y_t^4$, are central
to the phase structure we uncover: without this term competing against
the positive $\lambda^2$ feedback, there would be no finite critical
threshold $k_{\rm crit}$ at all.

The one-loop beta function for the Higgs quartic coupling, written in the
GUT-normalized convention for $g_1$ and with $\beta_\lambda =
d\lambda/d\ln\mu^2$, is
\begin{align}
\beta_\lambda^{(1)} &= \frac{1}{16\pi^2}\Big[12\lambda^2 + 6\lambda y_t^2
- 3y_t^4 - \tfrac{9}{2}\lambda g_2^2 - \tfrac{9}{10}\lambda g_1^2
\nonumber\\
&\quad + \tfrac{9}{16}g_2^4 + \tfrac{27}{400}g_1^4
+ \tfrac{9}{40}g_2^2 g_1^2\Big],
\label{eq:beta}
\end{align}
supplemented by the two- and three-loop contributions
$\beta_\lambda^{(2)}$ and $\beta_\lambda^{(3)}$ given explicitly in
Ref.~\cite{Buttazzo2013}. The crucial competition in
Eq.~(\ref{eq:beta}) is between the positive $12\lambda^2$ term
(self-reinforcing scalar feedback) and the negative $-3y_t^4$ term
(destabilizing top Yukawa contribution). Under the enhancement
$\lambda \to k\lambda$, the scalar term scales as $12k^2\lambda^2$, while
the Yukawa term $-3y_t^4$ is unchanged at the initial scale since the
enhancement applies only to $\lambda$, not to $y_t$. This asymmetry
creates the threshold behavior at $k_{\rm crit}$.

At the electroweak scale with SM values
$\lambda_{\rm SM} \approx 0.126$ and $y_t \approx 0.94$, the negative
top contribution dominates,
\begin{equation}
\beta_\lambda^{\rm SM} \propto 12(0.126)^2 - 3(0.94)^4
\approx 0.19 - 2.35 < 0,
\end{equation}
driving $\lambda$ toward instability. However, $y_t$ decreases at high
energies due to its own negative beta function. Around
$\mu \sim 10^{14}$~GeV where $y_t \approx 0.5$,
\begin{equation}
\beta_\lambda^{\rm enhanced} \propto 12k^2(0.126)^2 - 3(0.5)^4
\approx 0.19k^2 - 0.19.
\end{equation}
For $k > 1$, positive scalar feedback eventually overwhelms the
Yukawa contribution at high energies. The precise critical threshold
requires the full three-loop coupled evolution, which we implement
numerically. We have verified that the one-loop gauge and Yukawa beta
functions used in the numerical integration carry the standard SM
coefficients, $b_1 = 41/10$, $b_2 = -19/6$, and $b_3 = -7$.

\subsection{Numerical Implementation}

Starting from initial conditions at $\mu = M_t$,
\begin{align}
\lambda(M_t) &= k \times 0.126, \\
y_t(M_t) &= 0.937,\\
g_1(M_t) &= 0.461, \quad g_2(M_t) = 0.648, \quad g_3(M_t) = 1.167,
\end{align}
we integrate the full coupled system of RG equations, simultaneously
evolving $\lambda$, $y_t$, $g_1$, $g_2$, and $g_3$ with their complete
three-loop beta functions, to $\mu \sim 10^{19}$~GeV using adaptive
step-size methods in Maple. We track three diagnostics: (i) the
zero-crossing scale $\Lambda_I$ where $\lambda(\Lambda_I) = 0$ if it
occurs, (ii) the UV pole scale $\Lambda_{\rm UV}$ where
$\lambda \to \infty$, and (iii) the perturbativity scale
$\Lambda_{\rm pert}$ where $\lambda \sim 1$.

\section{Results}
\label{sec:results}

\subsection{Phase Structure and Critical Threshold}

\begin{figure}[h]
\centering
\includegraphics[width=0.7\textwidth]{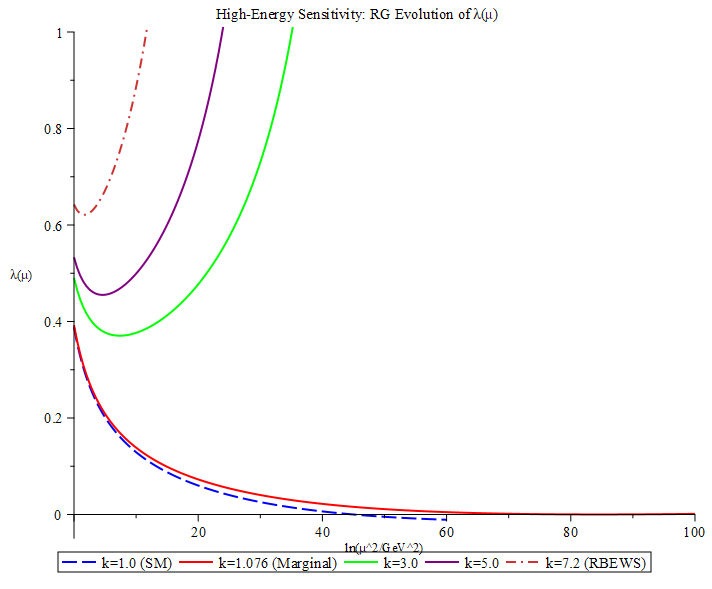}
\caption{High-energy sensitivity of the Higgs quartic coupling to
initial enhancement. The horizontal axis shows
$\ln(\mu^2/{\rm GeV}^2)$ and the vertical axis shows $\lambda(\mu)$.
The Standard Model ($k=1.0$) exhibits metastability with $\lambda \to 0$
near $\mu \approx 10^{10}$~GeV. The critical case
($k \approx 1.076$) represents marginal stability with $\lambda$
approaching but not crossing zero at high scales. Enhanced scenarios
($k=3.0$, $5.0$, $7.2$) show absolute stability with runaway growth
toward UV poles whose scale decreases rapidly with $k$, reaching
$\sim 10^5$~GeV for $k = 7.2$. All curves are obtained from the complete
three-loop SM RG equations including full Yukawa and gauge coupling
evolution.}
\label{fig:sensitivity}
\end{figure}

Figure~\ref{fig:sensitivity} displays the RG evolution of $\lambda(\mu)$
for representative enhancement factors. The results reveal three distinct
regimes with sharp transitions.

\textit{Region I: Metastable Vacuum ($k < k_{\rm crit}$).} For enhancement
factors below the critical threshold, the theory exhibits SM-like
behavior. The coupling $\lambda(\mu)$ decreases with increasing energy,
eventually crossing zero at the instability scale $\Lambda_I$. For
$k=1.0$ (SM), $\Lambda_I \approx 10^{10}$~GeV. As $k$ approaches
$k_{\rm crit}$ from below, $\Lambda_I$ increases dramatically, the
quantitative manifestation of near-criticality.

\textit{Region II: Marginal Stability ($k \approx k_{\rm crit}$).} At the
critical enhancement $k_{\rm crit} \approx 1.076$, corresponding to
$\lambda(M_t) \approx 0.135$, the coupling exhibits marginal behavior. It
decreases initially but asymptotes toward $\lambda \to 0^+$ at high
energies without crossing zero. This represents the phase boundary
separating metastable and absolutely stable regimes, and it coincides
with the conventional absolute-stability boundary expressed in
$\lambda(M_t)$.

\textit{Region III: Absolute Stability with UV Poles ($k > k_{\rm crit}$).}
Above the critical threshold, the theory transforms qualitatively. The
coupling remains positive at all scales and exhibits runaway growth
toward a Landau pole. The positive $\lambda^2$ feedback overwhelms the
negative Yukawa contribution at high energies, driving
$\beta_\lambda > 0$. The UV pole scale falls rapidly with increasing
enhancement, by many orders of magnitude between $k = 2$ and $k = 7.2$,
as quantified below.

Table~\ref{tab:results} quantifies the complete phase structure.

\begin{table*}[t]
\centering
\caption{High-energy behavior as a function of the enhancement factor $k$
in the corrected three-loop analysis. The zero-crossing scale $\Lambda_I$
indicates vacuum instability when present. The UV pole scale
$\Lambda_{\rm UV}$ marks effective field theory breakdown. The critical
threshold is $k_{\rm crit} \approx 1.076$ with $\lambda(M_t) \approx
0.135$, reproducing the conventional absolute-stability boundary.}
\label{tab:results}
\begin{ruledtabular}
\begin{tabular}{ccccc}
$k$ & $\lambda(M_t)$ & $\Lambda_I$ (GeV) & $\Lambda_{\rm UV}$ (GeV)
& Phase \\
\hline
1.00  & 0.126 & $\sim 10^{10}$ & \ldots             & Metastable \\
1.02  & 0.129 & $\sim 10^{11}$ & \ldots             & Metastable \\
1.04  & 0.131 & $\sim 10^{13}$ & \ldots             & Metastable \\
1.06  & 0.134 & $\sim 10^{16}$ & \ldots             & Metastable \\
1.076 & 0.135 & Never          & $\gtrsim 10^{19}$  & Marginal   \\
1.50  & 0.189 & Never          & $\gtrsim 10^{19}$  & Stable     \\
2.00  & 0.252 & Never          & $\sim 10^{18}$     & Stable     \\
3.00  & 0.378 & Never          & $\sim 10^{10}$     & Stable     \\
5.00  & 0.630 & Never          & $\sim 10^{6}$      & Stable     \\
7.20  & 0.907 & Never          & $\sim 10^{5}$      & Stable     \\
\end{tabular}
\end{ruledtabular}
\end{table*}

\subsection{Sensitivity Near the Critical Point}
\label{sec:sensitivity}

The transition at $k_{\rm crit} \approx 1.076$ exhibits pronounced
sensitivity to the scalar coupling. As $k$ approaches $k_{\rm crit}$ from
below, the coupling $\lambda(\mu)$ approaches zero with a near-vanishing
slope, and the zero-crossing scale $\Lambda_I$ becomes parametrically
sensitive to the initial condition, rising by many orders of magnitude
across a narrow interval in $k$. This corresponds to a logarithmic
susceptibility
\begin{equation}
\chi_\lambda \;\equiv\;
\frac{d\ln\Lambda_I}{d\ln k}\bigg|_{k \to k_{\rm crit}^-}
\;\sim\; \mathcal{O}(10^3),
\label{eq:susceptibility}
\end{equation}
which diverges as $k \to k_{\rm crit}^-$. No other sector of the
Standard Model exhibits a comparable response of a high-energy scale to a
low-energy parameter. For comparison, the analogous logarithmic
susceptibility of the QCD confinement scale $\Lambda_{\rm QCD}$ to the
strong coupling $\alpha_s(M_Z)$ is
$d\ln\Lambda_{\rm QCD}/d\ln\alpha_s \sim 1/(b_0\alpha_s) \sim 5$,
two to three orders of magnitude smaller than $\chi_\lambda$ near
criticality.

The structural origin of this susceptibility is the asymmetric scaling
of the two competing terms in $\beta_\lambda$ under the enhancement. The
positive scalar feedback term $12\lambda^2$ scales as $k^2$ at the
matching scale, while the destabilizing $-3y_t^4$ Yukawa term is unchanged
because the enhancement applies to $\lambda$ only. This asymmetry,
combined with the slow logarithmic running of $y_t$ between $M_t$ and
$\sim 10^{14}$~GeV, places the zero-crossing condition
$\beta_\lambda = 0$ at a scale that depends exponentially on the initial
value of $\lambda$.

We note explicitly that this sensitivity, expressed in $\lambda(M_t)$, is
the same near-criticality quantified by the Higgs-mass sensitivity of
Ref.~\cite{Buttazzo2013}, viewed through the coupling variable rather than
the mass variable. The boundary value $\lambda(M_t) \approx 0.135$ is
common to both descriptions; the distinction is conceptual, between a
BSM enhancement at fixed $M_h$ and a parameter-space variation of $M_h$
within the SM, as developed in Sec.~\ref{sec:intro}. The near-coincidence
of $k_{\rm crit}$ with the conventional bound is therefore not an
accident but the same physics in two parameterizations, and serves as a
validation of the implementation.

The phenomenological relevance of $\chi_\lambda \sim 10^3$ is immediate.
ATLAS measurements of the Higgs trilinear self-coupling constrain the
deviation parameter $\kappa_\lambda = \lambda_{hhh}/\lambda_{hhh}^{\rm
SM}$ at the level of a few units, with HL-LHC projections reaching
$|\kappa_\lambda - 1| \lesssim 0.5$ at $3\sigma$. For matching-scale BSM
modifications, the trilinear coupling tracks the quartic enhancement,
$\kappa_\lambda \approx k$ at tree level, with calculable loop-induced
corrections. The steep response of $\Lambda_I$ to a deviation in
$\kappa_\lambda$ converts modest experimental precision on the Higgs
self-coupling into qualitative information about the high-scale fate of
the theory.

\subsection{Radiative Symmetry Breaking Scenario}

With $k \approx 7.2$, the RBEWS benchmark~\cite{Steele2013} places the
theory deep in Region~III, with a matching-scale coupling
$\lambda(M_t) \approx 0.9$ that already approaches the perturbativity
limit at the weak scale. The vacuum is absolutely stable, but the
positive $\lambda^2$ feedback is so strong that $\lambda$ grows rapidly:
perturbative validity is lost by $\Lambda_{\rm pert} \sim 10^4$--$10^5$~GeV
and a Landau pole develops at $\Lambda_{\rm UV} \sim 10^{5}$~GeV. This is
a striking departure from the naive expectation that the breakdown scale
of an enhanced scenario lies near the grand-unified scale. For
$\lambda(M_t) \approx 0.9$, the breakdown of the perturbative effective
field theory occurs in the multi-TeV to PeV range, not near $M_{\rm GUT}$.

The physical implication is that a radiatively generated quartic of the
size predicted by Steele and Wang requires new strong dynamics, or a
compositeness scale, at $\sim 10^5$~GeV rather than at the unification
scale. This is a sharp and, in principle, testable consequence: the
RBEWS mechanism with large $e_{125}$ does not naturally embed in
high-scale unification through the running of $\lambda$ alone, but
instead predicts low-scale strong dynamics. We regard the location of
$\Lambda_{\rm UV}$ as an indicative scale for the onset of strong
coupling rather than a precision prediction, since perturbation theory is
not under control once $\lambda \sim 1$.

The SO(10) realization of Ref.~\cite{Chishtie2026PLB} produces a
matching-scale enhancement $k(M_t) \approx 6.49$, hence
$\lambda(M_t) \approx 0.8$, which lies in the same large-enhancement
regime. The perturbative breakdown scale implied by the SM running of
such a coupling is similarly low, and we note that the interpretation of
the high-scale behavior in that construction should be revisited in light
of the present corrected pole scales; the GUT-scale embedding there rests
on the matching physics at $M_{\rm GUT}$ rather than on the perturbative
running of $\lambda$ up to that scale.

\section{Discussion}
\label{sec:discussion}

\subsection{Interpretation of the UV Pole}

The Landau pole at $\Lambda_{\rm UV}$ should be viewed as signaling EFT
breakdown rather than as a pathology of the calculation. Above this scale
the SM ceases to be a valid effective description and new physics must
enter. For moderate enhancements just above $k_{\rm crit}$, the pole lies
near or above the Planck scale and the perturbative description is valid
over essentially the entire desert. For large enhancements the pole
scale collapses: by $k = 3$ it has fallen to $\sim 10^{10}$~GeV, and by
$k = 7.2$ to $\sim 10^5$~GeV. The breakdown scale is therefore a strongly
decreasing function of the enhancement, and the position of
$\Lambda_{\rm UV}$ identifies the scale at which a UV completion must
enter, not a domain where the SM description remains valid.

Several UV completions are natural at the relevant scale. Compositeness
scenarios where the Higgs emerges as a bound state of more fundamental
constituents~\cite{Kaplan1984} are well-motivated when the coupling is
large at low scales. Extended scalar sectors with additional singlets or
doublets can modify RG flow and soften or remove the pole through mixing
effects~\cite{Elias2003,Hiller2022,Hiller2024}. Asymptotic safety
proposals posit non-perturbative UV fixed points that regulate the Landau
pole~\cite{Shaposhnikov2009}. For the large-enhancement regime, the low
value of $\Lambda_{\rm UV}$ points specifically toward strong dynamics or
compositeness in the multi-TeV to PeV range rather than toward
unification-scale physics.

\subsection{Vacuum Stability Resolution}

Enhanced couplings provide a resolution to the SM metastability problem.
In the SM with $k=1.0$, the vacuum is metastable but long-lived, and the
125~GeV Higgs mass lies uncomfortably close to the critical value
separating stability and instability. For enhanced scenarios with
$k > k_{\rm crit}$, the vacuum is absolutely stable, with the electroweak
minimum representing the global minimum at all scales below
$\Lambda_{\rm UV}$. The cost of this resolution, in the
large-enhancement regime, is a low perturbative breakdown scale, which
must be addressed by an appropriate UV completion.

\subsection{Standard Model Baseline, Matching-Scale Dependence, and
its Limitations}
\label{sec:baseline}

Our sensitivity analysis characterizes the phase structure of scalar
enhancement at the electroweak matching scale, with $k_{\rm crit}
\approx 1.076$. The qualitative classification into metastable, marginal,
and stable phases reflects the SM renormalization group flow, and the
threshold expressed in $\lambda(M_t)$ is calculable a priori from the SM
beta functions.

A BSM mechanism may instead introduce a threshold correction to
$\lambda$ at a higher matching scale $\mu_{\rm match}$. The relative
enhancement at that scale,
\begin{equation}
k(\mu_{\rm match}) \;\equiv\;
\frac{\lambda_{\rm BSM}(\mu_{\rm match})}{\lambda_{\rm SM}(\mu_{\rm match})}
\;=\; 1 \,+\,
\frac{\Delta\lambda(\mu_{\rm match})}{\lambda_{\rm SM}(\mu_{\rm match})},
\label{eq:k_match}
\end{equation}
together with the corresponding threshold $k_{\rm crit}(\mu_{\rm match})$,
governs whether the trajectory is stabilized. We caution, however, that
$k(\mu_{\rm match})$ is not a universal model-building criterion and must
be interpreted with care. Two limitations are essential. First, the ratio
in Eq.~(\ref{eq:k_match}) is ill-conditioned at scales where
$\lambda_{\rm SM}(\mu)$ runs close to zero: there a small absolute
threshold correction $\Delta\lambda$ produces a numerically large
$k(\mu_{\rm match})$, but this large ratio reflects the smallness of the
denominator rather than a large physical effect, and the mapping to the
stability phase must be made through the absolute coupling
$\lambda_{\rm BSM}(\mu_{\rm match})$ rather than the ratio. Second, in a
genuine BSM model the new particles modify the beta functions above
$\mu_{\rm match}$, so the running is no longer the SM running and the
threshold shifts by model-dependent amounts. The SM-RG scan should
therefore be read as a baseline estimate of the size of the threshold
correction required under restrictive assumptions, not as a universal
phase diagram valid for arbitrary completions.

With these caveats, the SM-RG threshold provides a useful leading-order
reference. BSM corrections arise from three calculable sources. Additional
particles modify the beta functions above their mass thresholds: a
vectorlike fermion doublet with Yukawa $y_{\rm VL}$ contributes
$\delta\beta_\lambda \sim -N_c y_{\rm VL}^4/(16\pi^2)$, shifting the
threshold upward, while a scalar singlet with portal coupling $\eta$
contributes $+\eta^2/(16\pi^2)$, shifting it downward. Threshold
corrections at the BSM scale alter the matching condition by
$\mathcal{O}(\alpha_{\rm BSM}/4\pi)$. And if the UV completion involves
strong dynamics, perturbative running above $\Lambda_{\rm UV}$ ceases to
apply. The systematic analyses of Hiller et
al.~\cite{Hiller2022,Hiller2024} provide concrete instances: their
Planck-safe vacuum stability regions in singlet and vectorlike portal
extensions correspond, at the relevant matching scale, to effective
enhancements consistent with Region~III of Table~\ref{tab:results}, but
with model-specific beta-function modifications that the SM-RG baseline
does not capture.

\subsection{Phenomenological Implications}

Observable consequences arise at both collider and cosmological
frontiers. The Higgs trilinear coupling scales as
$\lambda_3 \propto \lambda$, directly proportional to the enhancement
factor. The di-Higgs production cross section $\sigma(gg \to HH)$ depends
on $\lambda^2$ with interference effects, potentially yielding factors of
$2$--$5$ enhancement for $k = 2$--$3$, observable at the High-Luminosity
LHC~\cite{Papaefstathiou2013}. As demonstrated in
Refs.~\cite{Hiller2022,Hiller2024}, the Higgs trilinear and quartic
couplings in BSM scenarios that stabilize the vacuum can deviate
significantly from SM predictions.

For the large-enhancement RBEWS regime, the prediction of strong dynamics
at $\sim 10^5$~GeV is the dominant phenomenological signature. While
direct production at such scales is beyond near-term colliders, the
associated large deviations in $\kappa_\lambda$ and the modified
high-scale running are accessible through precision Higgs measurements.
For RBEWS with $k \sim 7$, perturbative calculations of the phase
transition become unreliable, requiring non-perturbative methods such as
lattice simulations or functional renormalization group approaches.

\subsection{Validity, Limitations, and the EFT Perspective}
\label{sec:validity}

Our analysis employs SM RG equations with $\lambda(M_t)$ as a free
initial condition, providing a conservative and transparent baseline.
Several limitations deserve explicit discussion.

\textit{Beta function modifications.} Any BSM mechanism generating the
enhancement $k$ will introduce new particles that modify the SM beta
functions below their mass thresholds. The universal $12\lambda^2$ term
in $\beta_\lambda$ and the competing $-3y_t^4$ Yukawa contribution, the
two ingredients that generate $k_{\rm crit}$, are robust features of any
weakly-coupled scalar sector at one loop. For perturbative BSM spectra
with $M_{\rm BSM} \gtrsim 1$~TeV, corrections to $k_{\rm crit}$ are of
order $\mathcal{O}(10\%)$. For strongly-coupled completions, perturbative
beta functions cease to apply above the compositeness scale, which for
the large-enhancement regime is precisely the low scale where the Landau
pole appears.

\textit{Meaning of $\lambda$ at high energies.} The Landau pole marks the
scale at which the SM ceases to be a valid EFT. For $k = 7.2$,
perturbativity is already lost at $\Lambda_{\rm pert} \sim 10^4$--$10^5$~GeV,
which we identify as the practical boundary of the perturbative EFT
description. The pole position $\Lambda_{\rm UV}$ should therefore be
interpreted as an indicative scale for the onset of strong dynamics,
not a precision prediction beyond that boundary.

\textit{Scale dependence of the enhancement factor.} The RBEWS ratio
$e_{125} = \lambda_{\rm RBEWS}(v)/\lambda_{\rm CSB}(v) \approx 7.2$ is
defined at the vev scale in the CW scheme. Its value at $M_t$ in
$\overline{\rm MS}$ differs due to RG running and scheme conversion. We
use $k = 7.2$ as a benchmark; all central results, including
$k_{\rm crit}$, the phase structure, and the steep near-threshold
sensitivity, are independent of this specific value.

Despite these caveats, the central conclusions are secure. The transition
from metastable to absolutely stable vacuum at $k_{\rm crit} \approx
1.076$ is dictated by the competition between $\lambda^2$ and $y_t^4$
contributions in the beta function, and reproduces the conventional
absolute-stability boundary when expressed in $\lambda(M_t)$. The
emergence of UV poles for $k > k_{\rm crit}$, and the rapid collapse of
the pole scale with increasing $k$, follow from positive $\lambda^2$
feedback once $y_t$ has run down sufficiently.

\section{Conclusions}
\label{sec:conclusions}

We have analyzed Standard Model vacuum stability as a function of an
enhancement of the Higgs quartic coupling at the electroweak matching
scale, parameterized by $k = \lambda_{\rm enhanced}(M_t)/
\lambda_{\rm SM}(M_t)$ at fixed observed Higgs mass. This coupling scan is
a distinct physical question from the conventional Higgs-mass scan, though
the two share the same SM renormalization group flow.

Our three-loop analysis identifies a critical threshold
$k_{\rm crit} \approx 1.076$, an enhancement of roughly $7.6\%$,
separating metastability from absolute stability. The corresponding
matching-scale coupling $\lambda(M_t) \approx 0.135$ reproduces the
established three-loop absolute-stability boundary, an internal
consistency check that the present implementation agrees with the
conventional analysis when expressed in the common variable. The
instability scale responds steeply to $k$ near the threshold, a
logarithmic susceptibility $\chi_\lambda \sim \mathcal{O}(10^3)$ that
exceeds the analogous susceptibility of $\Lambda_{\rm QCD}$ to
$\alpha_s(M_Z)$ by two to three orders of magnitude, identifying the
Higgs sector as a sensitive probe of high-scale physics.

For $k > k_{\rm crit}$ the vacuum is absolutely stable and $\lambda$
develops an ultraviolet Landau pole whose scale falls rapidly with
increasing enhancement, from near the Planck scale for moderate $k$ to
$\sim 10^5$~GeV for the radiative symmetry breaking benchmark
$k \approx 7.2$. This collapse of the breakdown scale is the central
quantitative result of the corrected analysis. It implies that radiative
electroweak symmetry breaking with a large enhancement does not connect
to grand-unified-scale physics through the running of $\lambda$, but
rather predicts new strong dynamics or compositeness in the multi-TeV to
PeV range. This is a sharp, falsifiable consequence of the
large-enhancement scenario.

The phase structure provides a baseline for BSM model-building. We
caution that the relative enhancement $k(\mu_{\rm match})$ at a high
matching scale is not a universal criterion: it is ill-conditioned where
$\lambda_{\rm SM}$ runs near zero, and a genuine BSM completion modifies
the beta functions above the matching scale. The SM-RG threshold should
be read as a leading-order reference against which model-specific
corrections, calculable from a given particle spectrum, are measured.
This baseline is complementary to the systematic portal extension program
of Refs.~\cite{Hiller2022,Hiller2024}, which determines which crossing
points are phenomenologically consistent with precision electroweak and
collider constraints.

Detection of percent-level enhancements in the Higgs self-coupling at
future colliders would indicate an absolutely stable vacuum, while the
large enhancements associated with radiative symmetry breaking point to
strong dynamics at scales far below the unification scale, accessible in
principle through precision Higgs measurements and their implications for
the high-scale running.

\begin{acknowledgments}
F.A.C.\ and S.H.\ acknowledge and thank D.G.\ McKeon, T.G.\ Steele,
and Z.W.\ Wang for insightful discussions on RBEWS and SM scalar
coupling sensitivity. We thank G.\ Hiller, T.\ H\"ohne, D.F.\ Litim,
and T.\ Steudtner for drawing our attention to their work on Higgs
vacuum stability portals~\cite{Hiller2022,Hiller2024}, which provided
valuable context for the present analysis.
\end{acknowledgments}


\begin{thebibliography}{99}
\section*{References}
\bibitem{Aad2012}
G.~Aad et al.\ [ATLAS Collaboration],
Phys.\ Lett.\ B \textbf{716}, 1 (2012).

\bibitem{Chatrchyan2012}
S.~Chatrchyan et al.\ [CMS Collaboration],
Phys.\ Lett.\ B \textbf{716}, 30 (2012).

\bibitem{Buttazzo2013}
D.~Buttazzo, G.~Degrassi, P.P.~Giardino, G.F.~Giudice, F.~Sala,
A.~Salvio, and A.~Strumia,
JHEP \textbf{12}, 089 (2013) [arXiv:1307.3536].

\bibitem{Degrassi2012}
G.~Degrassi, S.~Di~Vita, J.~Elias-Miro, J.R.~Espinosa,
G.F.~Giudice, G.~Isidori, and A.~Strumia,
JHEP \textbf{08}, 098 (2012) [arXiv:1205.6497].

\bibitem{Hiller2022}
G.~Hiller, T.~H\"ohne, D.F.~Litim, and T.~Steudtner,
Phys.\ Rev.\ D \textbf{106}, 115004 (2022) [arXiv:2207.07737].

\bibitem{Hiller2024}
G.~Hiller, T.~H\"ohne, D.F.~Litim, and T.~Steudtner,
Phys.\ Rev.\ D \textbf{110}, 115017 (2024) [arXiv:2401.08811].

\bibitem{Coleman1973}
S.R.~Coleman and E.J.~Weinberg,
Phys.\ Rev.\ D \textbf{7}, 1888 (1973).

\bibitem{Steele2013}
T.G.~Steele and Z.-W.~Wang,
Phys.\ Rev.\ Lett.\ \textbf{110}, 151601 (2013).

\bibitem{Chishtie2026PLB}
F.A.~Chishtie,
Phys.\ Lett.\ B \textbf{876}, 140429 (2026).

\bibitem{Kaplan1984}
D.B.~Kaplan and H.~Georgi,
Phys.\ Lett.\ B \textbf{136}, 183 (1984).

\bibitem{Elias2003}
V.~Elias, R.B.~Mann, D.G.C.~McKeon, and T.G.~Steele,
Phys.\ Rev.\ Lett.\ \textbf{91}, 251601 (2003).

\bibitem{Shaposhnikov2009}
M.~Shaposhnikov and C.~Wetterich,
Phys.\ Lett.\ B \textbf{683}, 196 (2009).

\bibitem{Randall1999}
L.~Randall and R.~Sundrum,
Phys.\ Rev.\ Lett.\ \textbf{83}, 3370 (1999).

\bibitem{Contino2006}
R.~Contino, Y.~Nomura, and A.~Pomarol,
Nucl.\ Phys.\ B \textbf{671}, 148 (2003).

\bibitem{Branco2012}
G.C.~Branco et al.,
Phys.\ Rept.\ \textbf{516}, 1 (2012).

\bibitem{Wetterich2017}
C.~Wetterich,
Phys.\ Lett.\ B \textbf{773}, 6 (2017).

\bibitem{Meissner2007}
K.A.~Meissner and H.~Nicolai,
Phys.\ Lett.\ B \textbf{648}, 312 (2007).

\bibitem{Papaefstathiou2013}
A.~Papaefstathiou, L.L.~Yang, and J.~Zurita,
Phys.\ Rev.\ D \textbf{87}, 011301 (2013).

\bibitem{Caprini2020}
C.~Caprini et al.,
JCAP \textbf{03}, 024 (2020).

\end{thebibliography}
\end{document}